\documentstyle[fleqn,12pt]{article}

\def\gord{$ \raisebox{-.3ex}{$\stackrel{>}{_{\sim}}$} $}
\def\pj{\hspace{-.26cm}}
\def\mum{$\mu/M$}
\def\ms{$m_{\sigma}$}
\def\thalf{{\textstyle{\frac{1}{2}}}}
\def\tquar{{\textstyle{\frac{1}{4}}}}

\newcommand{\vmg}[1]{\mbox{\boldmath$#1$}}
\def\fpj{\hspace{-.7cm}}
\def\tpj{\hspace{-.3mm}}
\def\toneth{{\textstyle{\frac{1}{3}}}}
\begin{document}
\title{Neutron Stars and Nuclei in the Modified Relativistic
Hartree Approximation}
\author{M. Prakash$^{a,c}$, P. J. Ellis$^{b,d}$, E. K. Heide$^b$
and S. Rudaz$^b$\\[.5cm]
{\small $^a$Theoretical Physics Institute, University of Minnesota,
Minneapolis, MN 55455}\\[-.3cm]
{\small $^b$School of Physics and Astronomy, University of Minnesota,
Minneapolis, MN 55455}\\[-.3cm]
{\small $^c$Physics Dept., State University of New York at Stony Brook,
Stony Brook, NY 11794}\\[-.3cm]
{\small $^d$Institute for Nuclear Theory, University of Washington, Seattle,
WA 98195}}
\date{~}
\thispagestyle{empty}
\maketitle
\begin{center}
{\small\bf Abstract}
\vskip-.5cm
\end{center}
{\small We have examined the properties of neutron-rich matter and finite
nuclei in the
modified relativistic Hartree approximation for several values of the
renormalization scale, $\mu$, around the standard choice of $\mu$ equal to the
nucleon mass $M$.  Observed neutron star masses do not effectively constrain
the value of $\mu$. However for finite nuclei the value \mum=0.79, suggested by
nuclear matter data, provides a good account of the bulk properties with a
sigma mass of about 600 MeV.  This value of \mum ~renders the effective three
and four body scalar self-couplings to be zero at 60\% of equilibrium nuclear
matter density, rather than in the vacuum.  We have also found that the
matter part of the exchange diagram has little impact on
the bulk properties of neutron stars.}
\thispagestyle{empty}
\vskip-23cm
\phantom{1234}\hfill NUC-MINN-93/7-T\\
\phantom{1234}\hfill UMN-TH-1103/92\\
\phantom{1234}\hfill SUNY-NTG-92-21\\
\phantom{1234}\hfill MARCH, 1993
\newpage
\section{Introduction}

The archetypal Walecka model is a point-particle, relativistic field theory
in which nucleons interact via the exchange of $\sigma$- and $\omega$-mesons.
If systems with non-zero isospin are studied the exchange of $\rho$-mesons
also gives substantial contributions. This type of model has been widely
employed to study matter both at normal and at high density {\it e.g.}
\cite{sew}. The Lagrangian takes the form
\begin{eqnarray}
\hspace{-1cm}\cal L&\pj=&\pj\bar N\left(i\gamma^{\mu}\partial_{\mu}-g_{\omega}
\gamma^{\mu}\omega_{\mu}-\frac{g_{\rho}}{2}\gamma^{\mu}{\bf b}_{\mu}\cdot
\vmg \tau +\frac{f_{\rho}}{4M}\sigma^{\mu\nu}\partial_{\nu}{\bf b}_{\mu}
\cdot\vmg \tau -M+g_{\sigma}\sigma\right)N\nonumber\\
&&-\tquar F_{\mu\nu}F^{\mu\nu}+\thalf m^2_{\omega}\omega_{\mu}\omega^{\mu}
-\tquar {\bf B}_{\mu\nu}\cdot{\bf B}^{\mu\nu}+\thalf m^2_{\rho}{\bf b}_{\mu}
\cdot{\bf b}^{\mu}\nonumber\\
&&\qquad\qquad+\thalf\partial_{\mu}\sigma\partial^{\mu}\sigma
-\thalf m^2_{\sigma}\sigma^2\; ,
\end{eqnarray}
where
$F_{\mu\nu}=\partial_{\mu}\omega_{\nu}-\partial_{\nu}\omega_{\mu}$,
${\bf B}_{\mu\nu}=\partial_{\mu}{\bf b}_{\nu}-\partial_{\nu}{\bf b}_{\mu}$ and
$\sigma^{\mu\nu}=\thalf i[\gamma^{\mu},\gamma^{\nu}]$.
We have allowed for both vector and tensor coupling of the $\rho$ field,
${\bf b}_{\mu}$, to the
nucleon (the latter is only employed in evaluating the exchange or Fock
diagram). We note that the presence of the $\rho$ meson renders the
Lagrangian non-renormalizable, although
a renormalizable model can be constructed \cite{sew}.

In the mean field approximation, where the fields are replaced by their
classical expectation value, the energy density for an infinite system is
\begin{eqnarray}
{\cal E}_{\rm MFA} &\pj=&\pj \frac {C_\omega^2n^2}{2M^2}
+ \frac{M^2}{2C_{\sigma}^2}(M^*-M)^2+\frac{C_{\rho}^2}{8M^2}
(n_N-n_P)^2\nonumber\\
&&+ \frac {1}{\pi^2} \int\limits_0^{k_{FN}} dk\,k^2 E^*
+ \frac {1}{\pi^2} \int\limits_0^{k_{FP}} dk\,k^2 E^*\;.
\end{eqnarray}
Here  the nucleon effective mass, $M^*=M-g_{\sigma}\sigma$,
$E^*=\sqrt{k^2+M^{*2}}$ and
$C_i^2=(g_iM/m_i)^2$, where $i$ denotes the meson species.
The subscripts $N$ and $P$ on the Fermi momentum $k_F$
and density $n$ distinguish the neutron and proton values. The total density
$n=n_N+n_P$, thus the neutron excess parameter
$\alpha=\frac{N-Z}{A}=\frac{n_N-n_P}{n}$. Normal nuclear matter
corresponds to $\alpha=0$, but here we shall be interested in neutron star
matter for which $\alpha\sim1$. Also matter with $\alpha\sim\toneth$
is of interest for stellar collapse leading to supernova explosions.

In the relativistic Hartree approximation (RHA), where the one loop diagrams
are taken into account, the energy density is the sum of the mean field
contribution and a one-loop vacuum correction term
${\cal E}_{\rm RHA} = {\cal E}_{\rm MFA} +\Delta{\cal E}$.
The one--loop vacuum
correction term, caused by a shift in the single  particle energies,
receives contributions from the $\sigma$-interaction only. It contains
divergences and the necessary renormalization introduces a scale, $\mu$.
Explicitly \cite{hr},
\begin{equation}
\Delta{\cal E}(M^*)=-\frac{1}{4\pi^2}\left(M^{*4}\ln\frac{M^*}{\mu}
+\sum\limits_{n=0}^4f_nM^{4-n}(M-M^*)^n\right).
\end{equation}
In order to specify the finite parameters $f_n$, we
define $\lambda_m(M^*)=\frac{d^m\Delta{\cal E}(M^*)}{dM^{*m}}$ and require
$\lambda_0(M)$ and $\lambda_1(M)$ to be zero so that the vacuum energy
is zero and corresponds to a minimum. We can also set $\lambda_2(M^*)=0$
at $M^*=M$ because different choices simply correspond to modifying the
parameter $C^2_{\sigma}$ which is fitted to
equilibrium nuclear matter. The remaining parameters are specified
by requiring that $\lambda_{3}(M^*)$ and $\lambda_{4}(M^*)$, the effective
three and four body couplings, are zero at some
scale and for simplicity we choose the same scale $M^*=\mu$ in both cases. This
yields \cite{hr}
\begin{eqnarray}
f_0=\ln(\mu/M)\;,&&f_1=1-4\ln(\mu/M)\;,\nonumber\\
f_2=-{\textstyle\frac{7}{2}}+6\ln(\mu/M)\;,&&f_3={\textstyle\frac{25}{3}}-4
\mu/M\;,\hspace{.7cm}f_4=-{\textstyle\frac{25}{12}}\;.
\end{eqnarray}
The vacuum correction term $\Delta{\cal E}$ can then be written in the form
\begin{eqnarray}
\Delta{\cal E}&\pj=&\pj-\frac{M}{\pi^2}\left(1-\frac{\mu}{M}+
\ln\frac{\mu}{M}\right)(M-M^*)^3\nonumber\\
&&\qquad\qquad+\frac{1}{4\pi^2}\ln\frac{\mu}{M}\ (M-M^*)^4+
\Delta{\cal E}_{\rm CW}\;.
\end{eqnarray}
For the standard choice \cite{sew}, $\mu=M$, the first and second terms on the
right vanish and one is left with the Chin-Walecka \cite{chin} result
\begin{eqnarray}
\Delta{\cal E}_{\rm CW}&\pj=&\pj-\frac{1}{4\pi^2}\biggl[M^{*4}
\ln\frac{M^*}{M}+M^3(M-M^*)-{\textstyle \frac{7}{2}}M^2(M-M^*)^2\nonumber\\
&&\qquad\qquad+{\textstyle \frac{13}{3}}
M(M-M^*)^3-{\textstyle \frac{25}{12}}(M-M^*)^4\biggr]\;.
\end{eqnarray}
\vskip-.25cm
Note that eq. (6) is designed to have no $\sigma^3$ or $\sigma^4$
contribution, such terms arise for $\mu\ne M$ and are shown explicitly in
eq. (5). We refer to this approach as the modified relativistic Hartree
approximation (MRHA).

In principle the results of an exact treatment would be independant of the
renormalization scale, however this will not be the case when a simple
approximation such as the MRHA is employed. Therefore
the renormalization scale should be chosen on physical grounds and
we have previously found \cite{hr,rehp} that in nuclear matter the choice
$\mu\ne M$ gives a better account of the second derivative of the
energy/particle
at saturation (the compression modulus) and also the third derivative (the
skewness or anharmonicity).
Our purpose here is to test the MRHA further
by applying it to finite nuclei and neutron stars.
Nuclei are expected to provide a more stringent test, in particular
they are much more sensitive to the value of the nucleon
effective mass than infinite systems. The value of
$M^*$ in the interior of a heavy nucleus should approximate that found
for equilibrium nuclear matter, which varies significantly with the
renormalization scale. We will examine the predictions of the MRHA for the
spherically-symmetric, doubly-closed-shell
nuclei $^{16}$O, $^{40}$Ca and $^{208}$Pb. In the case of neutron stars
we are dealing with dense asymmetric matter. Apart from the obvious question
of whether the MRHA supports neutron stars of the observed mass, we ask whether
such global properties provide a significant constraint on the nuclear equation
of state around the saturation density. It is also of interest to ask whether
the proton fractions
are such that the direct Urca process, {\it i.e.} neutrino emission from a
single nucleon, is possible. If so, this will provide the dominant cooling
mechanism \cite{prak} since the standard modified Urca mechanism involves two
nucleons.

Exchange terms are a fundamental feature of nuclear physics, but they are
not included in the MRHA.
An additional goal of this work is to study the contribution of the
lowest-order, two-loop exchange diagram to nuclear and neutron-rich
matter. The vacuum contributions are
large in magnitude, however it has been found that they
become small if form factors are included to reflect the composite
nature of the particles \cite{pek}. We therefore
consider just the matter, or Fermi sea, contributions which are much
less affected by form factors.
This is a tractable calculation which should allow us to get a qualitative
indication of the importance of exchange effects in dense matter.

The organization of the paper is  as follows. Section 2 deals with
the MRHA; we start with a brief discussion of nuclear matter in subsec. 2.1,
turn to asymmetric matter in subsec. 2.2, neutron stars in
subsec. 2.3. and, finally, finite nuclei in subsec. 2.4. In sec. 3 we
investigate the
effect of the matter part of the lowest-order exchange diagram for
infinite systems. Our conclusions are given in sec. 4.

\section{The Modified RHA}
\subsection{Nuclear Matter}
For a given value of \mum\ the parameters $C^2_{\omega}$ and $C^2_{\sigma}$ are
fitted to the equilibrium nuclear matter properties. We take a binding
energy/particle of 16 MeV. In the interior region of heavy nuclei the measured
charge density indicates an equilibrium matter density of $n_0=0.16\pm0.01$
fm$^{-3}$; we choose the central value of 0.16 fm$^{-3}$.
The parameter $C^2_{\rho}$ is fitted to the coefficient of the symmetry
energy, $a_{\rm symm}=\frac{1}{2n}\frac{\partial^2{\cal E}}
{\partial\alpha^2}\left|_{\alpha=0}\right.$, which
is taken to be 30 MeV. The
resulting parameters are listed in table 1 for a range of values of
$\mu/M$. The omega coupling constants, $C_{\omega}^2$,
agree qualitatively with the experimental result \cite{omeg} of $103\pm36$
and are within the cited errors for $\mu/M$= 0.73, 1.00 and 1.25. Table 1
also gives the calculated compression moduli,
$K=k_F^2\frac{d^2}{dk_F^2}\left(\frac{E}{A}\right)$.
However, Pearson \cite{pear} has made the point that $K$ is not uniquely
determined by the limited amount of accurate breathing mode data available
(see also Shlomo and Youngblood \cite{shlomo}). He determines a linear
correlation between $K$ and the coefficient of the Coulomb term,
$K_{\rm coul}Z^2/A^{\frac{4}{3}}$, in the leptodermous expansion. This
coefficient can be conveniently written
\begin{equation}
K_{\rm coul}=-\frac{3}{5}\frac{e^2}{r_0}\left[2+\frac{S}{K}\right]\;,
\end{equation}
where $r_0$ is the radius parameter and $S$ is the skewness or
anharmonicity parameter defined by
$S=k_F^3\frac{d^3}{dk_F^3}\left(\frac{E}{A}\right)$.
One then has a linear correlation between
$S/K$ and $K$ as shown by the shaded band in fig. 1 (the box indicates the
point obtained by Sharma \cite{sh}, however $\chi^2$ changes little over
the shaded area). The points give results \cite{rehp}
obtained in the MRHA for various values of the renormalization scale, $\mu$.
The MRHA results lie within the error band for values of the renormalization
scale $\approx$ 0.79 and 1.25. The other values of
$\mu/M$ in table 1 lie off the shaded band and are included in our discussion
for comparison purposes.
The value 0.73 has the attractive feature that it is the self-consistent
point where $\mu=M^*_{\rm sat}$.

\subsection{Asymmetric Matter}

We show in fig. 2 the binding energy/particle for nuclear matter and for pure
neutron matter as a function of density; the full curves are obtained in the
present MRHA approximation. At a given density the difference in
energy between the pairs of curves is roughly independent of $\mu$,
varying by less than 10\%. As regards
the stiffness of the equation of state, it is often assumed that
this can be deduced qualitatively from the compression modulus. We see,
however, that it is not always a reliable indicator. Thus at high density the
equation of state for $\mu=0.73M$ is noticably stiffer than for $\mu=1.25M$,
yet the compression modulus is smaller. A quantitative measure of the high
density    behavior of the equation of state is provided by the adiabaticity
index, $\Gamma=(d\ln P/d\ln n)$. At densities $n>2.5n_0$ the value of $\Gamma$
for both nuclear and neutron matter lies in the range 2--3, with little
dependence on the renormalization scale apart from the case $\mu=0.73M$. Here
the pressure increases more rapidly with density and $\Gamma$ is significantly
larger than in the other cases.

As we add neutrons to nuclear matter, i.e., as $\alpha$ increases from 0,
the equilibrium binding energy is reduced and the corresponding density
is reduced. We have verified that this can be accurately described by the
expression \cite{neutm}
\begin{equation}
n_{\rm sat}(\alpha)=n_{\rm nm}(1-C\alpha^2)\;.
\end{equation}
The constant $C$ is listed in Table 2. Further one can define the
compression modulus at the saturation density of the asymmetric system, i.e.,
the isobaric incompressibility $K_{\rm iso}$, and to second order in $\alpha$
this can be written
\begin{equation}
K_{\rm iso}(\alpha)=K\left[1+(A-B)\alpha^2\right]\equiv K\left[1-
\widetilde{A}\alpha^2\right]\;.
\end{equation}
Here the coefficient $A$ gives the asymmetry dependence of $K_{\rm iso}$
at nuclear matter density, while $B$ allows for the change of equilibrium
density in the neutron-rich system. These parameters, along with the
combination
$\widetilde{A}$ are given in table 2. Again we remark that this quadratic
form provides a very accurate approximation, excepting only
values of $\alpha$ close to the point where the system becomes unbound.
The values of $C$ and $\widetilde{A}$ are smallest for \mum=1 and show a factor
of 3 variation with renormalization scale.  Other calculations \cite{neutm}
show a rather wide variation in the predicted values of $C$. However
$\widetilde{A}$ is close to 1.6 in the chiral $\sigma$-model and it usually
lies in the range 1.3--2.2 in non-relativistic calculations, although larger
values are to be found. These are comparable to the values in table 2.

\subsection{Stellar Matter}

We have calculated the properties of neutron stars using the hydrostatic
equilibrium equations of Tolman, and Oppenheimer and Volkov \cite{tov}.  For $n
> 0.08$ fm $^{-3}$,  the MRHA equation of state is used.    For densities
$0.001 < n<0.08$ fm $^{-3}$, we employ the  equation of state of Negele and
Vautherin \cite{nv}, while for very low densities $(n < 0.001$ fm $^{-3})$, we
use the Baym-Pethick-Sutherland \cite{bps} equation of state.  The calculation
is carried out for charge neutral matter in beta equilibrium.  The condition
for beta equilibrium is
\begin{equation}
\mu_e = \mu_N - \mu_P = - \frac{\partial E (n,x)}{\partial x}\;,
\end{equation}
where $\mu_i$, $i=N,P$ and $e$, are the chemical potentials of the fermions,
$x=Z/A$ and $E(n,x)$ is the energy per baryon.  Muons will be present when
$\mu_e\geq m_\mu = 105.7$ MeV, which is generally the case for $n \geq n_0$.
In the presence of muons, the proton fractions are determined by imposing the
charge neutrality condition $n_e+n_\mu=n_P$ and the energy conservation
condition $\mu_e=\mu_\mu$.  The lepton contributions to the energy density
and pressure are given by Fermi gas expressions (the electromagnetic
interactions of the leptons give negligible contributions).

The measured mass \cite{mass} of 4U0900-40, which is $(1.85\pm0.3)M_{\odot}$,
may provide  a limit for the maximum mass of a neutron star. The most
accurate determination \cite{puls}, for one component of the binary pulsar
PSR $1913+16$, is $(1.44\pm0.01)M_{\odot}$. The maximum masses in table 3 are
consistent with these limits, although since hyperons are
expected to reduce the maximum mass \cite{glen,kapol} the case with $\mu=1.5M$
can probably be excluded and the value for $\mu=1.25M$ appears dangerously
low. Note that $M_{\rm max}$ is not a monotonic function of $K$. Thus observed
neutron star masses cannot be used to effectively constrain the compression
modulus alone.
Such properties of {\it nuclear matter} are not uniquely determined
by the gross features of neutron stars \cite{pal}, which constrain the softness
of the neutron-rich equation of state. In table 3 we also give the
radius of the neutron star and the ratio of the central density to equilibrium
nuclear matter density, the latter for a maximum mass star and a
$1.44M_{\odot}$ star. The qualitative trends follow the high density stiffness
of the equation of state.

We turn now to estimate the maximum Keplerian frequency of
rotation from the formula \cite{hz}
\begin{equation}
\Omega_K\simeq0.77\left(\frac{M_{\rm max}}{M_{\odot}}\right)^{\!\frac{1}{2}}
\left(\frac{R_{\rm max}}{10\ {\rm km}}\right)^{\!-\frac{3}{2}}\ 10^4{\rm
s}^{-1}
\;,
\end{equation}
where $M_{\rm max}$ and $R_{\rm max}$ are the maximum mass and corresponding
radius for the spherical non-rotating star. The empirical relation (11)
reproduces the results of more exact calculations \cite{rot,latt}. It is
interesting to observe
that changes in the renormalization scale serve to increase the Keplerian
rotation frequency and give values which are on the upper end of those found
with relativistic equations of state \cite{latt}.

Recently, it has been pointed out~\cite{prak} that the direct Urca processes
\begin{equation}
n \rightarrow p+e^-+{\overline \nu}_e\quad, \quad
p + e^- \rightarrow n + \nu_e
\end{equation}
may constitute the principal avenues through which the late-time rapid  cooling
of neutron stars occurs.  For these processes to occur in matter in which the
only baryons are nucleons, momentum conservation requires that the magnitude of
the electron concentration $x_e$ exceeds a value~\cite{prak}
\begin{equation}
|x_e|^{1/3} \geq |(1-x)^{1/3} - x^{1/3}|\,,
\label{momc}
\end{equation}
where $x=Z/A$ is the proton fraction.  When leptons ($e^-$ and $\mu^-$) are the
only source of negatively charged particles, charge neutrality and
eq.~(\ref{momc}) stipulate  that $x$ exceed a value in the range $0.11-0.15$.
In the models considered above, the symmetry energy rises sufficiently rapidly
with density so that   the calculated proton fractions are large enough for the
direct Urca cooling mechanism to be operative.

Since the calculated maximum neutron star masses exceed the observed values for
all but very large values of \mum,  we conclude that the gross properties of
neutron stars do not effectively constrain the renormalization scale \mum.

\subsection{Finite Nuclei}
\subsubsection{Formalism}

For finite nuclei the Coulomb interaction is needed so that the Lagrangian
(1) is supplemented by the photon Lagrangian
\begin{equation}
{\cal L}_{ph}=-\tquar f_{\mu\nu}f^{\mu\nu}-e\bar N\thalf(1+\tau_3)
\gamma^{\mu}A_{\mu}N\;.
\end{equation}
Here $A_{\mu}$ is the Maxwell field and $f_{\mu\nu}$ is the electromagnetic
field strength tensor. It is straightforward \cite{sew,hs} to obtain from
${\cal L}+{\cal L}_{ph}$ the equations of motion in the mean field
approximation. The vacuum is taken into account by means of the local density
approximation and the leading correction terms of the derivative expansion
\cite{perry}. The latter give small, although not negligible corrections
\cite{was}.
The derivative expansion appears to be rapidly convergent \cite{wako,blun}
so that the lowest order corrections should suffice.

The corrections of the derivative expansion can be obtained by evaluating
the one-loop polarization contribution to the scalar and vector meson
propagators \cite{furn}. The question arises as to where these propagators
should be renormalized,
at $M^*=M$ or at $M^*=\mu$. If in the former case one chooses $\lambda_2(M)=0$
and in the latter $\lambda_2(\mu)=0$, where $\lambda_2$ is defined in
sec. 1, then changing the renormalization point for the scalar meson is simply
equivalent to varying \ms, which is a free parameter anyway. For the
$\omega$-meson it is reasonable to require that, in vacuum, the pole of the
propagator lies at the known physical mass. Then changes in the
renormalization point
will not affect the predictions, but rather the interpretation of the
parameters. Therefore it is sufficient to take the derivative contributions
in standard form \cite{perry} so that the complete scalar density is
\begin{eqnarray}
\rho_s(r)&\pj=&\pj\sum\limits_{\alpha}^{\rm occ}\frac{(2j_{\alpha}+1)}{4\pi
r^2}
\left[G_{\alpha}(r)^2-F_{\alpha}(r)^2\right]+\frac{d\Delta{\cal E}}{dM^*}
\nonumber\\
&\pj-&\pj\frac{1}{4\pi^2}\hspace{-1mm}\left[2g_{\sigma}\ln\frac{M^*}{M}
\left(\frac{d^2\sigma}{dr^2}+\frac{2}{r}\frac{d\sigma}{dr}\right)\hspace{-1mm}
-\frac{g_{\sigma}^2}{M^*}\left(\frac{d\sigma}{dr}\right)^{\!\!2}\hspace{-2mm}
-\frac{2g_{\omega}^2}{3M^*}\left(\hspace{-1mm}\frac{d\omega_0}{dr}\hspace{-1mm}
\right)^{\!\!2}\right]\hspace{-1mm},
\end{eqnarray}
where $G$ and $F$ are the components of the Dirac spinors for the occupied
states and $M^*$ is the spatially dependant effective mass,
$M^*(r)=M-g_{\sigma}\sigma(r)$. Only the timelike component of $\omega_{\mu}$
is non-zero for the spherically symmetric nuclei we consider.
The full vector density is
\begin{eqnarray}
\rho_B(r)&\pj=&\pj\sum\limits_{\alpha}^{\rm occ}\frac{(2j_{\alpha}+1)}{4\pi
r^2}
\left[G_{\alpha}(r)^2+F_{\alpha}(r)^2\right]\nonumber\\
&&\pj-\frac{1}{3\pi^2}\left[g_{\omega}\ln\frac{M^*}{M}\left(\frac{d^2\omega_0}
{dr^2}+\frac{2}{r}\frac{d\omega_0}{dr}\right)-\frac{g_{\sigma}g_{\omega}}{M^*}
\left(\frac{d\sigma}{dr}\frac{d\omega_0}{dr}\right)\right]\;,
\end{eqnarray}
and the contribution to the total energy from the vacuum is given by
\begin{equation}
\Delta E=4\pi\hspace{-2mm}\int\limits_0^{\infty}\hspace{-1mm}r^2dr\hspace{-1mm}
\left[\Delta{\cal E}(M^*)-\frac{g_{\sigma}^2}{4\pi^2}\ln\frac{M^*}{M}
\left(\frac{d\sigma}{dr}\right)^{\!\!2}\hspace{-2mm}+\frac{g_{\omega}^2}{6\pi^2}
\ln\frac{M^*}{M}\left(\frac{d\omega_0}{dr}\right)^{\!\!2}\right]\hspace{-1mm},
\end{equation}
where $M^*=M^*(r)$.

\subsubsection{Results}

We now consider the predictions of the MRHA for finite nuclei\footnote{This is
also under investigation by Blunden \cite{peter} with similar results where
comparison can be made.}, specifically
the doubly closed-shell nuclei $^{16}$O, $^{40}$Ca and $^{208}$Pb.
In addition to the parameters of table 1, the meson masses are required.
We use the physical masses for the $\rho$- and $\omega$-mesons; the value
of $m_{\sigma}$, which is not a priori known, is discussed below. A pure vector
$\rho N$ coupling is employed in these calculations.
As we have remarked, the vacuum is treated by means of the
local density approximation supplemented by the leading terms of
the derivative expansion. The effect of the latter are fairly small, giving
a decrease in the binding energy, an increase in the radius for a fixed
$\sigma$ mass and very little change in the single particle energies.
If $m_{\sigma}$ is adjusted so that similar radii are obtained
with and without the derivative terms, the net result is a small increase
in the binding energy when the derivative terms are included \cite{was}.
We shall present results which include the contributions of the derivative
terms.

In table 4 we summarize the bulk properties of the nuclei as a function of
\mum\ for a fixed value of \ms= 550 MeV. The theoretical values of the binding
energy/particle include a correction for the c.m. kinetic energy \cite{neg}.
The charge densities, and therefore the radii quoted, are corrected for the
finite size of the proton \cite{hs} and for c.m. effects \cite{tas}. In table
4 we also give the separation between the neutron $2f$ levels in $^{208}$Pb
so as to indicate the order of magnitude of the spin-orbit splittings.
In studying table 4 it must be borne in mind that while $C^2_{\sigma}$
is fixed from nuclear matter, the value of \ms\ can be regarded as a free
parameter. Decreasing \ms\ has the effect of increasing $r_{ch}$ and
decreasing $\frac{BE}{A}$ and vice versa \cite{hs}. Thus for \mum=0.73
the charge radius of Pb is too large, while those of O and Ca are too
small so that we shall not be able to obtain a reasonable account of
these nuclei by varying \ms. For \mum=1.5,
the deviations of the radius from experiment are quite disparate for Ca
and Pb and this cannot be resolved by adjusting \ms. Further in this case the
spin-orbit splittings are much too small-- the value in table 4
is less than a third of the experimental figure. This is due to
the large value of the effective mass at saturation in nuclear matter
(0.86). As is well known \cite{sew} smaller effective masses
$\sim0.6$ yield better values for the spin-orbit splittings and this is evident
from comparison of tables 1 and 4. The value \mum=1.25 also yields quite small
spin-orbit splittings due to a large value of $M^*_{\rm sat}$, although,
for comparison purposes, we will consider it along with the values 0.79 and
1.0 in subsequent calculations.

The results obtained by varying \ms\ so as to approximately fit the
radius of Ca and, if possible, also Pb  are given in table 5. The binding
energies for \mum=0.79 are in strikingly good agreement with the data
for a reasonable value of \ms=600 MeV. With \mum=1.25 the binding energies
are low and the required $\sigma$ mass is unreasonably small. The value
\mum=1.0 also yields rather low binding energies and further one cannot
obtain a good account of the charge radii of Ca and Pb simultaneously.
We should emphasize at this juncture that our results are obtained by
fitting to a nuclear matter saturation density, $n_0=0.16$ fm$^{-3}$.
A number of other authors, e.g. \cite{sew,hs,was,fox}, have used the value
0.15 fm$^{-3}$ which represents a reduction in $k_F$ of 2\%. This has the
effect
of increasing the nuclear radius by approximately 2\% so that
a larger \ms\ can be used and this increases the binding energies. For
example, Wasson \cite{was} takes 550 MeV (\mum=1.0) which gives binding
energies
1--2 MeV short of the experimental value and in fact for \mum=0.79 the nuclei
are overbound with this saturation density. In both cases close agreement
with the observed radii for all three nuclei cannot be obtained.

Adopting the values of \ms\ in table 5 which are appropriate to $n_0=0.16$
fm$^{-3}$, we obtain the single particle energies of $^{208}$Pb
which are shown in figs. 3 and 4 for neutrons and
protons respectively. We indicate both occupied and unoccupied experimental
states and their theoretical counterparts, if bound, for levels
near the Fermi energy. The shell closure and rough level ordering are
reproduced quite well, however the occupied neutron
(proton) levels are too strongly (weakly) bound which could be improved by
increasing the $\rho$ coupling \cite{hs}. For example an increase in
$g_{\rho}^2$ of 40\%, corresponding to a symmetry energy of 35 MeV, would
shift the occupied single particle levels in the right direction by about
1 MeV; the unoccupied levels are shifted by a somewhat smaller amount.

The spin orbit splittings that we find are too small; on average they are 80,
60
and 30\% of the experimental values for \mum=0.79, 1.0 and 1.25, respectively,
and the percentage becomes somewhat smaller for the lighter nuclei. For the
standard renormalization, Fox \cite{fox} has made the same point and noted that
the mean field results are close to, or a little larger than, the data (see
also, for example, the fits in ref. \cite{stray}). Finally we show in figs.
5--7 the comparison of the predicted charge distributions with the data
(experimental errors are not indicated since they are negligible except near
the center of the nuclei). In Pb for \mum=1.0 our central density is too high,
whereas Fox \cite{fox} achieves better agreement, presumably due to his choice
of $n_0=0.15$ fm$^{-3}$. We do however find a region of positive slope,
although it not as pronounced as the data; mean field calculations do better in
this respect. Our results for \mum=0.79 and 1.25 are significantly better, both
for radial distances of 2--5 fm and in the important tail region. For Ca,
\mum=1.0 clearly gives the best agreement with the data, although the other
cases are not unreasonable. In O the different theoretical  results are of
comparable quality.

Viewing the results for finite nuclei overall, the renormalization scale
\mum=0.79 gives the best representation of the data. The additional terms which
are introduced into the vacuum correction $\Delta{\cal E}$ were given in eq.
(5).  This value of \mum ~renders the  self-consistently determined effective
three and four body scalar self-couplings to be zero at a nuclear
density of $n=0.092$ fm$^{-3}$, which is roughly halfway between the
equilibrium point and the vacuum.   With \mum=0.79 the
coefficient of the induced $\sigma^4$ term is negative and this has long been
known to be necessary in mean field treatments which fit parameters of
$\sigma^3$ and $\sigma^4$ terms to nuclei \cite{stray,boup,gam}.

\section{Exchange Contributions for Infinite \protect \\ Systems}
\subsection{Formalism}

We turn now to assess the importance of exchange contributions from
 the two-loop diagram shown in fig. 8; here a solid
line represents a nucleon with propagator $S$ and a dashed line represents
a meson with propagator $D$. Along with the $\sigma$ and $\omega$ mesons we
shall include the $\rho$ since this is clearly necessary for neutron-rich
matter. The exchange diagram gives a contribution to the
energy density
\begin{equation}
{\cal E} = \frac{1}{2}\, \int \frac{d^4p_1}{(2\pi)^4} \frac{d^4p_2}{(2\pi)^4}\,
{\rm Tr}[S(p_1)S(p_2)]D(k) \, ,
\end{equation}
where $k=p_1-p_2$ and Lorentz indices have been suppressed.  We use the mean
field form of the propagators, treating the two-loop contribution as a
perturbation. The matter part of this contribution to the energy density
involves the basic integrals
\begin{eqnarray}
I(w,\xi_1,\xi_2)&\pj=&\pj\int\limits_1^{\xi_1}du\left(1-\frac{1}{u^2}
\right)J(w,u,\xi_2)\;,\nonumber\\
J(w,u,\xi_2)&\pj=&\pj\tquar\int\limits_1^{\xi_2}dv\left(1-\frac{1}{v^2}\right)
\ln\frac{(uv-1)^2+uvw}{(u-v)^2+uvw}\;.
\end{eqnarray}
For completeness, we first give the expressions \cite{chin} for $\sigma$- and
$\omega$-exchange. The energy density is the sum of neutron and proton
contributions
\begin{eqnarray}
{\cal E}_{\sigma}&\pj=&\pj e_{\sigma}\left(k_{FN}\right)
+e_{\sigma}\left(k_{FP}\right)\;,\nonumber\\
{\cal E}_{\omega}&\pj=&\pj e_{\omega}\left(k_{FN}\right)
+e_{\omega}\left(k_{FP}\right)\;.
\end{eqnarray}
Here
\begin{eqnarray}
e_{\sigma}\left(k_F\right)&\pj=&\pj\frac{g_{\sigma}^2}{64\pi^4}\left[
(k_FE_F^*-M^{*2}\ln\xi)^2+M^{*4}(4-w_{\sigma})I(w_{\sigma},\xi,\xi)\right]\,,
\nonumber\\
e_{\omega}\left(k_F\right)&\pj=&\pj\frac{g_{\omega}^2}{32\pi^4}\left[
(k_FE_F^*-M^{*2}\ln\xi)^2-M^{*4}(2+w_{\omega})I(w_{\omega},\xi,\xi)\right]\,,
\end{eqnarray}
where
\begin{eqnarray}
E_F^{*2}&=&k_F^2+M^{*2}, \qquad \qquad \xi=(k_F+E_F^*)/M^*, \nonumber \\
w_{\sigma}&=&(m_{\sigma}/M^*)^2\qquad {\rm and}
\qquad w_{\omega}=(m_{\omega}/M^*)^2. \nonumber
\end{eqnarray}
The energy density for $\rho$ exchange can be written
\begin{equation}
{\cal E}_{\rho}=e_{\rho}\left(k_{FN},k_{FN}\right)
+e_{\rho}\left(k_{FP},k_{FP}\right)
+4e_{\rho}\left(k_{FN},k_{FP}\right)\;.
\end{equation}
For the general case with both vector and tensor coupling each
$e_{\rho}$ is the sum of vector-vector, tensor-tensor and vector-tensor
contributions
\begin{equation}
e_{\rho}\left(k_{F1},k_{F2}\right)
=e_{\rho}^{\rm vv}\left(k_{F1},k_{F2}\right)
+e_{\rho}^{\rm tt}\left(k_{F1},k_{F2}\right)
+e_{\rho}^{\rm vt}\left(k_{F1},k_{F2}\right)\;.
\end{equation}
These contributions can be written in the form
\begin{eqnarray}
&&\fpj e_{\rho}^{\rm vv}\left(k_{F1},k_{F2}\right)=\left(\frac{g_{\rho}}
{2}\right)^{\!2}\!\!\frac{1}{32\pi^4}\Big[
(k_{F1}E_{F1}^*-M^{*2}\ln\xi_1)(k_{F2}E_{F2}^*-M^{*2}\ln\xi_2)\nonumber\\
&&\qquad\qquad\qquad\qquad\qquad\qquad-M^{*4}(2+w_{\rho})I(w_{\rho},\xi_1,\xi_2)
\Big]\;,\nonumber\\
&&\fpj e_{\rho}^{\rm tt}\left(k_{F1},k_{F2}\right)=\left(\frac{f_{\rho}}
{4M}\right)^{\!2}\frac{M^{*2}}{64\pi^4}\Big[(10+w_{\rho})
(k_{F1}E_{F1}^*-M^{*2}\ln\xi_1)\nonumber\\
&&\ \ \times(k_{F2}E_{F2}^*-M^{*2}\ln\xi_2)
-\frac{8k_{F1}^3k_{F2}^3}{9M^{*2}}-M^{*4}w_{\rho}(8+w_{\rho})
I(w_{\rho},\xi_1,\xi_2)\Big],\nonumber\\
&&\fpj e_{\rho}^{\rm vt}\left(k_{F1},k_{F2}\right)=\tpj\left(\frac{g_
{\rho}f_{\rho}}{8M}\right)\tpj\frac{3M^*}{16\pi^4}\Big[(k_{F1}E_{F1}^*\tpj-\tpj
M^{*2}\ln\xi_1)(k_{F2}E_{F2}^*\tpj-\tpj M^{*2}\ln\xi_2)\nonumber\\
&&\qquad\qquad\qquad\qquad\qquad\qquad-M^{*4}w_{\rho}I(w_{\rho},\xi_1,\xi_2)
\Big]\;,
\end{eqnarray}
where $w_{\rho}=(m_{\rho}/M^*)^2$ and the subscripts 1 and 2 on $E_F$ and
$\xi$ refer to the appropriate Fermi momenta. If the no-retardation
approximation is
made, {\it i.e.}, energy differences ($E_1^*-E_2^*$) are set to zero, this can
be put in the form given by Bouyssy {\it et al.} \cite{bou}. Note that they
remove the  (three-dimensional) delta function contribution from the
tensor-tensor piece, arguing that it should be suppressed by short-range
correlations.

Thus in this approximation the total energy density
${\cal E}_{\rm total}={\cal E}_{\rm RHA}+{\cal E}_{\rm exch}$, where
${\cal E}_{\rm exch}={\cal E}_{\sigma}+{\cal E}_{\omega}+{\cal E}_{\rho}$.
For a given density the value of $M^*$ is obtained from the usual minimization
condition $\partial{\cal E}_{\rm total}/\partial M^*=0$ and here both the
RHA and exchange contributions are taken into account. A word of caution
is in order regarding the high density behavior, $k_F\rightarrow\infty$.
Since in this limit ${\cal E}_{\rm MFA}\propto n^2$, while ${\cal E}_{\sigma}$
and ${\cal E}_{\omega}\propto n^{\frac{4}{3}}$ \cite{chin} these exchange
contributions are negligible. However the $\rho$ tensor coupling involves a
derivative which brings in an extra power of the momentum. This causes the
tensor-tensor contribution to dominate the asymptotic behavior so that
${\cal E}_{\rho}\propto-n^2$ and the total energy density
${\cal E}\propto(C^2_{\omega}-\frac{(3-\alpha^2)}{128}f^2_{\rho})n^2$ has the
possibility
of becoming negative for sufficiently large $f_{\rho}$. Furthermore the
equation for the effective mass $M^*$ will ultimately be dominated by the
$\rho$ vector-tensor and tensor-tensor contributions which will give rise to a
negative value of $M^*$ (for positive $f_{\rho}/g_{\rho}$). Such problems
arise in principle for the $\rho$, but only for densities
$\gord15n_0$ which are well beyond our range of interest and probably beyond
the range of applicability of  a simple baryon-meson picture.

The Landau effective mass, $M^*_L$, is defined in terms of the velocity at the
Fermi surface, $v_F$, and the single particle spectrum, $\epsilon(k)$,
according
to \cite{bc}
\begin{equation}
M^*_L=\frac{k_F}{v_F}\;,\ \ {\rm where}\ \ v_F=\frac{\partial\epsilon(k)}
{\partial k}\bigg|_{k=k_F}\;.
\end{equation}
The well known result $M^*_L=(k_F^2+M^{*2})^{\frac{1}{2}}$ is obtained in
the MRHA. Here we shall study the modification of $M^*_L$ induced by the
exchange diagram for the case of nuclear matter where the neutron and proton
Fermi momenta are the same. In order to perform the calculation we need
$\epsilon(k)$. This can
conveniently be obtained by taking the functional derivative of the energy
density with respect to the occupation probability,
$\delta{\cal E}/\delta n({\bf k})$, where, for nuclear matter,
$n({\bf k})=(4\pi^3)^{-1}\theta(k_F-|{\bf k}|)$. The contributions of scalar
and vector exchange to $\epsilon(k)$ have been given by Chin \cite{chin}.
The contribution of the $\rho$ can be written as the sum of vector-vector,
tensor-tensor and vector-tensor terms,
\begin{equation}
\epsilon_{\rho}(k)=\epsilon_{\rho}^{\rm vv}(k)+\epsilon_{\rho}^{\rm tt}(k)
+\epsilon_{\rho}^{\rm vt}(k)\;.
\end{equation}
For nuclear matter these terms are
\begin{eqnarray}
\epsilon_{\rho}^{\rm vv}(k)&\pj=&\pj\left(\frac{g_{\rho}}{2}\right)^{\!2}
\frac{3}{8\pi^2E_k^*}\left[k_FE_F^*-M^{*2}\ln\xi-(2+ w_{\rho})\frac{M^{*3}}{k}
J\left(w_{\rho},\xi_k,\xi\right)\right],\nonumber\\
\epsilon_{\rho}^{\rm tt}(k)&\pj=&\pj-\left(\frac{f_{\rho}}{4M}\right)^{\!2}
\frac{1}{8\pi^2E_k^*}\biggl[2k_F^3E_k^*-3M^{*2}(5+\thalf w_{\rho})
(k_FE_F^*-M^{*2}\ln\xi)\nonumber\\
&&\qquad\qquad\qquad+\frac{12M^{*5}}{k}w_{\rho}(1+{\textstyle\frac{1}{8}}
w_{\rho})J\left(w_{\rho},\xi_k,\xi\right)\biggr]\;,\nonumber\\
\epsilon_{\rho}^{\rm vt}(k)&\pj=&\pj\left(\frac{g_{\rho}f_{\rho}}{8M}\right)
\frac{9M^*}{4\pi^2E_k^*}\left[k_FE_F^*-M^{*2}\ln\xi-\frac{M^{*3}}{k}w_{\rho}
J\left(w_{\rho},\xi_k,\xi\right)\right],
\end{eqnarray}
where $E_k^{*2}=k^2+M^{*2},\ \xi_k=(k+E_k^*)/M^*$ and the other quantities
are defined above. The derivative of $\epsilon$ is then easily evaluated to
obtain the contribution to the Fermi velocity.

\subsection{Results}

When the exchange diagram is included it is necessary to refit nuclear matter
properties. In order to restrict
the number of parameters, we take a $\sigma$ mass of 550 MeV and
fix the $\rho$-nucleon couplings. For the vector
part we take $g^2_{\rho}/4\pi=2.4$ ($C_{\rho}^2=44.9$) from $\pi N$
scattering \cite{ho}, assuming $\rho$ universality. Reference \cite{ho}
indicates a tensor-vector ratio, $f_{\rho}/g_{\rho}$, of 6.6, which is at the
upper end of currently accepted values \cite{omeg}. We shall also consider the
smaller ratio of 3.7 derived from the isovector anomalous magnetic moment of
the nucleon in the vector dominance model; studies of nuclear systems in the
Hartree-Fock
approximation \cite{bou} favor the smaller value. The $\sigma$ and
$\omega$ coupling constants which result from fitting the saturation
properties are collected in table 6. Since the net effect of exchange is
repulsive, it is necessary to reduce the $\omega$ coupling and for \mum=1.5,
$C^2_{\omega}$ is
unreasonably small and, in fact, we were not able to obtain a satisfactory
fit with the stronger tensor coupling. We also show in table
6 the compression moduli which show some modest
changes from their previous values; for $\mu/M=0.79$ and 1.0, $K$ is reduced,
whereas in the other cases $K$ increases. In the last column we give
the coefficient of the symmetry energy, $a_{\rm symm}$. Here the exchange
contribution from the $\sigma$ ($\omega$) is positive (negative), while the
effect of the $\rho$ is quite small; the net contribution from exchange
is large, $\sim+$15 MeV, as is well known. Thus without the direct $\rho$
contribution $a_{\rm symm}\sim 30$ MeV and, adding in the $\rho$, one obtains
a value in the neighborhood of 40 MeV. Given that currently accepted values are
in the range 27--35 MeV \cite{mol,pea}, our results are somewhat high
particularly for $\mu/M<1$.

The effect of exchange on the binding energy/particle as a function of density
is shown in fig. 2. The dashed curves correspond to the weaker tensor $\rho$
coupling for which the effect of exchange is fairly small, although the
impact of the increased symmetry energy is noticeable for neutron matter.
The dash-dotted curves show the results for the stronger tensor coupling;
the reduced values of $C^2_{\omega}$ here lead to a softening of the equation
of state for densities beyond 0.43 fm$^{-3}$ in all cases.
The parameters $C$ and
$\widetilde A$ of eqs. (8) and (9) characterise the change in equilibrium
density and compression modulus with the neutron excess parameter $\alpha$.
The effect of exchange upon these quantities is case dependant, but typically
the values of $C$ in table 2 are increased by 10\% and the values of
$\widetilde A$ are increased by 20\%.

In table 7 we show the individual contributions of the MRHA and exchange to the
Fermi velocity (see eq. (22)). The exchange contribution arises from
$\sigma$ (negative) and $\omega$ and $\rho$ (positive) contributions, none of
which is insignificant. The total exchange contribution is  about 5\%
(25\%) for the weak (strong) tensor coupling.  In the weak case the net results
for the Landau effective mass are quite close to those given in table 1 for the
pure MRHA, whereas  a reduction of $\sim$10\% is obtained with the strong
tensor coupling. In neither case are the exchange contributions to the
effective mass sufficiently small that they can safely be ignored.

Turning to neutron stars, we find that properties given in table 3 are little
changed when exchange effects are included  using the weaker tensor $\rho$
coupling. With the stronger tensor coupling the equation of state is softer
which leads to a modest increase in the central densities and a small
decrease in the maximum masses. The predictions for \mum=0.79 and 1.0
become very similar, in particular the maximum mass is close
to $2.26M_{\odot}$. The predictions for \mum=0.73 and 1.25 are also similar
and the maximum mass of $1.79M_{\odot}$ is probably a little low in view of
the expected effect of hyperons \cite{glen,kapol}. The cooling of neutron
stars by the direct Urca process is allowed in all cases, as was the case
without exchange.

Summarizing this section, we have examined the matter contribution arising from
the lowest order exchange diagram using the minimum number of mesons and found
quite modest effects for neutron-rich systems, even
with the strong tensor-vector ratio for the $\rho$ coupling. It is unlikely
that inclusion of, for example, the pion field would qualitatively change
this conclusion. Indeed rather small effects were obtained from the Fock terms
in (pure matter) Hartree-Fock calculations \cite{ww}
which included the pion as well as a number of baryon fields.

\section{Conclusions}

The main thrust of our work has been to examine the implications of different
choices of the renormalization scale in neutron-rich matter and finite nuclei,
following our previous work \cite{hr,rehp} with nuclear matter. Physical
results would be independant of this scale if we had embarked on a complete
renormalization program which accounted for all the important physical effects,
but, of course, we cannot claim that this is the case. Within the confines of
the relativistic Hartree approximation we have previously found \cite{rehp}
that nuclear matter favors values of \mum=0.79 and 1.25. The observed masses of
neutron stars do not distinguish different values of \mum; however, for finite
nuclei  we find that the phenomenology with \mum=0.79 is favored since it
provides rather reasonable agreement with the wide range of data considered. We
can compare with the mean field approach (ref. \cite{sew} and references
therein) which has the disadvantage that it excludes the vacuum which is an
inherent feature of relativistic field theories. Provided $\sigma^3$ and
$\sigma^4$ terms are included in the Lagrangian, this approach gives somewhat
better agreement for finite nuclei, particularly as regards the single particle
energies. However it must be borne in mind that non-relativistic calculations
\cite{all} show that correlations play an important role in determining the
spectral strength distributions in actual nuclei.

Changing the renormalization scale \mum\ from 1.0 to 0.79 can be viewed as
choosing the effective three and four body scalar self-couplings to be zero at
a point which correspond to 60\% of equilibrium nuclear matter density, rather
than in the vacuum. For \mum$< 1$, the coefficient of the $\sigma^4$ term is
negative and one can worry that the effective potential of the scalar field,
\begin{displaymath}
U_{\rm eff}(\sigma)=\thalf m^2_{\sigma}\sigma^2+\Delta{\cal E}(M-
g_{\sigma}\sigma)\;,
\end{displaymath}
\noindent has quite a shallow minimum \cite{rehp} at the vacuum value of
$\sigma=0$ for \mum=0.79. More troublesome is the fact that for
$\sigma\rightarrow-\infty$ the effective potential is unbounded from
below, $U_{\rm eff}\rightarrow-\infty$, irrespective of the choice
of the renormalization scale. This means that formally we are dealing with a
local minimum rather than an absolute minimum in the energy. Thus, while the
phenomenology is successful in explaining a large amount of data, unresolved
theoretical issues remain.

We thank C.J. Horowitz for giving us a copy of his computer code for finite
nuclei and for discussions pertaining thereto.
We acknowledge partial support from the Department of Energy under contracts
No. DE--FG02--87ER40328, DE--AC02--83ER40105 and DE--FG02--88ER40388. A grant
for computing time from the Minnesota Supercomputer Institute is gratefully
acknowledged.

\newpage
\centerline{\bf Table 1}
\centerline{MRHA coupling constants and saturation properties}
\begin{center}
\begin{tabular}{|cccccc|}\hline
${\displaystyle \frac{\mu}{M}}$&
${\displaystyle\frac{\strut M^*_{\rm sat}}{M}}$&$K$&$C^2_{\omega}$&
$C^2_{\sigma}$&$C^2_{\rho}$\\
&&(MeV)&&&\\ \hline
0.73&0.73&157&132.8&279.4&82.5\\
0.79&0.66&354&180.6&317.5&73.5\\
1.00&0.73&461&137.7&215.0&81.6\\
1.25&0.82&264&\phantom{1}78.6&178.6&90.8\\
1.50&0.86&189&\phantom{1}51.2&175.6&94.4\\ \hline
\end{tabular}
\end{center}
\vskip1cm
\centerline{\bf Table 2}
\centerline{Parameters specifying the saturation density and isobaric}
\centerline{compression modulus of neutron-rich matter}
\begin{center}
\begin{tabular}{|ccccc|}\hline
${\displaystyle \frac{\mu}{M}}$&
$C$&$A$&$B$&$\widetilde{A}$\\[1mm] \hline
0.73&1.62&3.36&5.45&2.09\\
0.79&0.76&1.56&4.45&2.89\\
1.00&0.54&1.05&2.16&1.10\\
1.25&0.91&1.75&2.96&1.21\\
1.50&1.25&2.39&3.64&1.25\\ \hline
\end{tabular}
\end{center}
\newpage
\centerline{\bf Table 3}
\centerline{Neutron star properties in the MRHA}
\begin{center}
\begin{tabular}{|ccccccc|}\hline
${\displaystyle \frac{\mu}{M}}$&$K$&
${\displaystyle\frac{\strut M_{\rm max}}{M_{\odot}}}$&$R$&
${\displaystyle \frac{n_c}{n_0}}$&${\displaystyle \frac{n_c}{n_0}}$&
$\Omega_K$\\
&(MeV)&&(km)&$(M_{\rm max})$&$(1.44M_{\odot})$&$(10^4\ {\rm s}^{-1})$\\ \hline
0.73&157&2.18&11.2&6.2&2.6&0.96\\
0.79&354&2.53&12.7&4.8&1.9&0.86\\
1.00&461&2.30&12.1&5.3&2.0&0.88\\
1.25&264&1.86&10.7&7.3&3.1&0.95\\
1.50&189&1.59&\phantom{1}9.9&9.1&4.5&0.99\\ \hline
\end{tabular}
\end{center}
\vskip.25cm
\begin{center}
{\bf Table 4}
\vskip 0.04in
Bulk properties of nuclei for $m_{\sigma}=550$ MeV
\vskip 0.04in
\begin{tabular}{|c|cc|cc|ccc|} \hline
&\multicolumn{2}{c|}{O}&\multicolumn{2}{c|}{Ca}&\multicolumn{3}{c|}{Pb}\\
$\frac{\mu}{M}$&$\frac{BE}{A}$&$r_{ch}$&$\frac{BE}{A}$&$r_{ch}$&
$\frac{BE}{A}$&$r_{ch}$&$2f\:\nu$ split\\
&(MeV)&fm&(MeV)&fm&(MeV)&fm&(MeV)\\ \hline
0.73&\phantom{1}9.42&2.66&\phantom{1}9.52&3.42&8.33&5.57&1.09\\
0.79&\phantom{1}7.52&2.80&\phantom{1}7.99&3.54&7.47&5.54&1.64\\
1.00&\phantom{1}6.08&2.64&\phantom{1}6.92&3.38&6.70&5.40&1.31\\
1.25&\phantom{1}9.16&2.51&\phantom{1}9.26&3.28&7.99&5.40&0.73\\
1.50&10.59&2.46&10.31&3.24&8.56&5.43&0.50\\ \hline
Expt.&\phantom{1}7.98&2.73&\phantom{1}8.45&3.48&7.86&5.50&1.77\\ \hline
\end{tabular}
\end{center}
\newpage
\begin{center}
{\bf Table 5}
\vskip 0.04in
Bulk properties of nuclei with $m_{\sigma}$ fitted
\vskip 0.04in
\begin{tabular}{|cc|cc|cc|cc|} \hline
&&\multicolumn{2}{|c|}{O}&\multicolumn{2}{c|}{Ca}&\multicolumn{2}{c|}{Pb}\\
$\frac{\mu}{M}$&$m_{\sigma}$&$\frac{BE}{A}$&$r_{ch}$&$\frac{BE}{A}$&$r_{ch}$&
$\frac{BE}{A}$&$r_{ch}$\\
&(MeV)&(MeV)&fm&(MeV)&fm&(MeV)&fm\\ \hline
0.79&600&8.13&2.74&8.47&3.49&7.73&5.51\\
1.00&450&4.03&2.79&5.35&3.49&5.82&5.45\\
1.25&350&5.53&2.77&6.65&3.48&6.63&5.50\\ \hline
\multicolumn{2}{|c|}{Experiment}&7.98&2.73&8.45&3.48&7.86&5.50\\ \hline
\end{tabular}
\end{center}
\vskip.25cm
\centerline{\bf Table 6}
\centerline{MRHA+ exchange coupling constants and saturation properties}
\begin{center}
\begin{tabular}{|cccccc|}\hline
${\displaystyle \frac{\mu}{M}}$&$K$&$C^2_{\omega}$&
$C^2_{\sigma}$&${\displaystyle\frac{f_{\rho}}{g_{\rho}}}$&
$a_{\rm symm}$\\[-2mm]
&(MeV)&&&&(MeV)\\\hline
0.73&191&\phantom{1}88.3&257.1&3.7&41.7\\[-2mm]
    &204&\phantom{1}57.9&212.0&6.6&38.6\\[-1mm]
0.79&332&128.9&283.6&3.7&42.2\\[-2mm]
    &274&\phantom{1}96.4&247.1&6.6&39.7\\[-1mm]
1.00&460&106.7&201.5&3.7&34.8\\[-2mm]
    &421&\phantom{1}80.9&186.6&6.6&34.4\\[-1mm]
1.25&275&\phantom{1}49.5&169.6&3.7&35.4\\[-2mm]
    &280&\phantom{1}26.3&149.5&6.6&34.6\\[-1mm]
1.50&198&\phantom{1}18.7&169.3&3.7&37.9\\ \hline
\end{tabular}
\end{center}
\newpage
\begin{center}
{\bf Table 7}
\vskip 0.04in
Landau effective masses in the MRHA+exchange approximation
\vskip 0.04in
\begin{tabular}{|c|ccc|ccc|} \hline
\mum&\multicolumn{3}{c|}{$f_{\rho}/g_{\rho}=3.7$}&\multicolumn{3}{c|}
{$f_{\rho}/g_{\rho}=6.6$}\\
&\multicolumn{2}{c}{Contribution to $v_F$}&$M^*_L/M$&\multicolumn{2}{c}
{Contribution to $v_F$}&$M^*_L/M$\\
&MRHA&exchange&&MRHA&exchange&\\ \hline
0.73&0.350&0.011&0.78&0.339&0.086&0.66\\
0.79&0.378&0.021&0.70&0.364&0.089&0.62\\
1.00&0.354&0.038&0.72&0.348&0.103&0.62\\
1.25&0.322&0.023&0.81&0.318&0.095&0.68\\
1.50&0.308&0.012&0.87&$\cdot$&$\cdot$&$\cdot$\\ \hline
\end{tabular}
\end{center}
\newpage
\newpage
\begin{center}
{\bf Figure Captions}
\end{center}

\noindent Fig. 1.\hspace{.3cm}The ratio of the third and second derivatives
of the nuclear matter equation of state at the equilibrium point, $S/K$,
versus the
compression modulus, $K$. The shaded band is obtained from the breathing mode
data \cite{pear} and the filled square is the value suggested by Sharma
\cite{sh}. The MRHA points are labelled by the value of the renormalization
scale, $\mu/M$.

\noindent Fig. 2.\hspace{.3cm}The binding energy/particle as a function of
density for neutron and nuclear matter with four values of the
renormalization scale. The full curves correspond to the MRHA and the
dashed (dash-dotted) curves correspond to MRHA + exchange with the weaker
(stronger) tensor $\rho$ coupling. The upper (lower) set of curves are for
neutron (nuclear) matter.

\noindent Fig. 3.\hspace{.3cm}Occupied and unoccupied neutron levels near the
Fermi energy in $^{208}$Pb. The experimental data are compared with
predictions for \mum=0.79, 1.0, 1.25.

\noindent Fig. 4.\hspace{.3cm}As for fig. 5, but for protons.

\noindent Fig. 5.\hspace{.3cm}Comparison of the experimental charge density
\cite{sick} for $^{16}$O with theoretical
predictions for \mum=0.79, 1.0, 1.25.

\noindent Fig. 6\hspace{.3cm}As for fig. 5, but for $^{40}$Ca. The data are
from ref. \cite{vri}.

\noindent Fig. 7\hspace{.3cm}As for fig. 6, but for $^{208}$Pb.

\noindent Fig. 8.\hspace{.3cm}The exchange or Fock diagram. In our
calculations the dashed line represents the exchange of a $\sigma$-,
$\omega$- or $\rho$-meson.
\end{document}